
\input harvmac

\let\preprint=0    


\def\PSfig#1#2#3#4#5{
  \nfig{#1}{#2}
  \if\preprint1
    \topinsert
      \dimen1=#3                     
      \divide\dimen1 by 100
      \multiply\dimen1 by #5
      \vskip \dimen1
      \count253=83                   
      \multiply\count253 #5
      \divide\count253 by 100
      \count254=100                  
      \advance\count254 by -#5
      \multiply\count254 by 254      
      \divide\count254 by 100
      \advance\count254 by -65       
      \dimen1=#3                     
      \advance\dimen1 by -11in
      \divide\dimen1 by 2
      \count255=\dimen1
      \divide\count255 by 100
      \multiply\count255 by \count253
      \dimen2=1bp
      \divide\count255 by \dimen2
      \advance\count255 by 18        
      \includegraphics{#4}
      \setbox1=\hbox{#1. #2}
      \ifdim \wd1 < 4.5in
         \centerline{#1. #2}
      \else
         \centerline{\vbox{\hsize 4.5in \baselineskip12pt \noindent #1. #2}}
      \fi
    \endinsert
  \fi
}


\def\refmark#1{${}^{\refs{#1}}$\ }
\def\footsym{*}\def\footsymbol{}\ftno=-2
\def\foot{\ifnum\ftno<\pageno\xdef\footsymbol{}\advance\ftno by1\relax
\ifnum\ftno=\pageno\if*\footsym\def\footsym{$^\dagger$}\else\def\footsym{*}\fi
\else\def\footsym{*}\fi\global\ftno=\pageno\fi
\xdef\footsymbol{\footsym\footsymbol}\footnote{\footsymbol}}

\def\ggg{{>}\!\!{>}\!\!{>}}

\def\mh{m_{\rm\scriptscriptstyle h}}
\def\mt{m_{\rm\scriptscriptstyle t}}
\def\d{{\rm d}}
\def\gy{g_{\rm y}}

\def\Vcl{V_{\rm cl}}
\def\Veff{V_{\rm eff}}
\def\Rc{R_{\rm c}}
\def\Ec{E_{\rm c}}
\def\SE{S_{\rm\scriptscriptstyle E}}

\lref\av{P. Arnold and S. Vokos, \sl Phys.\ Rev.\ \bf D44\rm, 3620 (1991).}
\lref\sher{M. Sher, \sl Phys.\ Rep.\ \bf 179\rm, 274 (1989).}
\lref\arnold{P. Arnold, \sl Phys.\ Rev.\ \bf D40\rm, 613 (1989).}
\lref\flores{R. Flores and M. Sher, \sl Phys.\ Rev.\ \bf D27\rm, 1679 (1982);
  M. Duncan, R. Philippe and M. Sher, \sl Phys.\ Rev.\ \bf 153B\rm,
    165 (1985); \sl Phys.\ Lett.\ \bf B209\rm, 543(E) (1988).}
\lref\greg{G. Anderson, \sl Phys.\ Lett.\ \bf B243\rm, 265 (1990).}
\lref\ellis{J. Ellis, A. Linde and M. Sher, \sl Phys.\ Lett.\ \bf B252\rm,
  203 (1990).}
\lref\hsu{S. Hsu, \sl Phys.\ Lett.\ \bf B261\rm, 81 (1991).}
\lref\brezin{E. Brezin and G. Parisi, \sl J. Stat.\ Phys.\ \bf 19\rm, 269
  (1978).}
\lref\everyone{
  I. Krive and A. Linde, \sl Nucl.\ Phys.\ \bf B42\rm, 873 (1979);
  N. Krasnikov, \sl Yad.\ Fiz.\ \bf 28\rm, 549 (1978);
  P.Q. Hung, \sl Phys.\ Rev.\ Lett.\ \bf 42\rm, 873 (1979);
  A. Anselm, \sl JETP Lett.\ \bf 29\rm, 590 (1979);
  N. Cabibbo, L. Maiani, A. Parisi and R. Petronzio,
    \sl Nucl.\ Phys.\ \bf B158\rm, 295 (1979).
}
\lref\twoloop{
  For a 2-loop calculation, see M. Lindner, M. Sher and H. Zaglauer,
  \sl Phys.\ Lett.\ \bf B228\rm, 139 (1989).
}

\Title{\vbox{
    \hbox{UW/PT-92-25}
  }}{\vbox{
    \centerline{A Review of the Instability of Hot Electroweak Theory}
    \centerline{and its Bounds on $\mh$ and ${\mt}^*$
  }}\footnote{}{${}^*$ Talk presented
            at the International Seminar Quarks `92,
            Zvenigorod, Russia, May 11--17, 1992.}}
\centerline{Peter Arnold}
\centerline{\sl Department of Physics, University of Washington,
    Seattle, WA  98195}

\vskip .5in
The electroweak vacuum need not be absolutely stable.  For certain top
and Higgs masses in the Minimal Standard Model, it is instead
metastable with a lifetime exceeding the present age of the Universe.
The decay of our vacuum may be nucleated at low temperature by
quantum tunneling or at high temperature in the early Universe
by thermal excitation.
I briefly review the constraints on top and Higgs masses from requiring that
the electroweak vacuum be sufficiently stable to have survived to the
present day.

\Date{December 1992}

\centerline{\bf A Review of the Instability of Hot Electroweak Theory}
\centerline{\bf and its Bounds on $\mh$ and $\mt$}

\baselineskip=14pt plus 1pt minus 1pt

\vskip 0.2in
\centerline{Peter B. Arnold}
\vskip 0.2in
\centerline{Department of Physics}
\centerline{University of Washington, Seattle, WA  98195}
\vskip 0.2in

\parskip=0pt
\abovedisplayskip=3pt
\belowdisplayskip=3pt

\sequentialequations

\newsec{Introduction}

Today I discuss a question of {\it life} and {\it death} importance---though,
admittedly, only to our descendants a billion times removed.
Is the electroweak vacuum absolutely
stable?  Or will it instead someday decay away, billions of years hence, and
wipe out the Universe as we know it?  In pursuing this question, we shall
be led to focus on a related question that should have concerned our
ancestors a billion times removed:
was the electroweak vacuum stable in the hot
Early Universe?  Investigating these concerns of the far-flung past and
far-flung future will eventually address a question of
relevance today: what are the best bounds on the top and Higgs masses
that can be derived from requiring
that the electroweak vacuum be sufficiently stable that it could have
persisted to the present day?  This discussion will all be in the
context of the Minimal Standard Model, and I shall rashly assume
that the standard single-Higgs-doublet model of electroweak symmetry
breaking happens to be a good effective theory of nature below some
large scale $\Lambda \ggg 1$ TeV.

Much of this discussion is a synopsis of Ref.~\av, which is work that
I did with Stamatis Vokos.
For a good review of much of the earlier work on the
subject, see Ref.~\sher.

To understand that the electroweak vacuum can be unstable, consider
the one-loop running of the Higgs self-coupling $\lambda$.  It is
of the form:
\eqn\brun{
  \beta_\lambda \equiv {\d\lambda\over\d(\ln M)}
  \sim a \lambda^2 + b g^2 - c \gy^4
}
where $a$, $b$ and $c$ are positive constants.  The three terms are due to the
contributions of scalar, vector, and fermion loops respectively, where
$g$ and $\gy$ are the gauge and Yukawa couplings.
The minus sign in front of the fermion loop contribution comes from Fermi
statistics.
Now suppose that the top mass is large compared to the W and Higgs masses.
Then $\gy^2$ is large compared to $g^2$ and $\lambda$ and the last term
dominates:
\eqn\brunf{
  \beta_\lambda \sim - c \gy^4 .
}
The running of $\lambda$ is shown qualitatively in fig.~1.
At large scales $M$, $\lambda$ runs negative.  This means that
the Higgs sector becomes {\it unstable} at very large mass scales
or very short distances.

\PSfig\figa{The running of $\lambda$ with scale $M$ for large top mass.
  The end-point on the left is fixed by the physical
  Higgs mass.}{9.2cm}{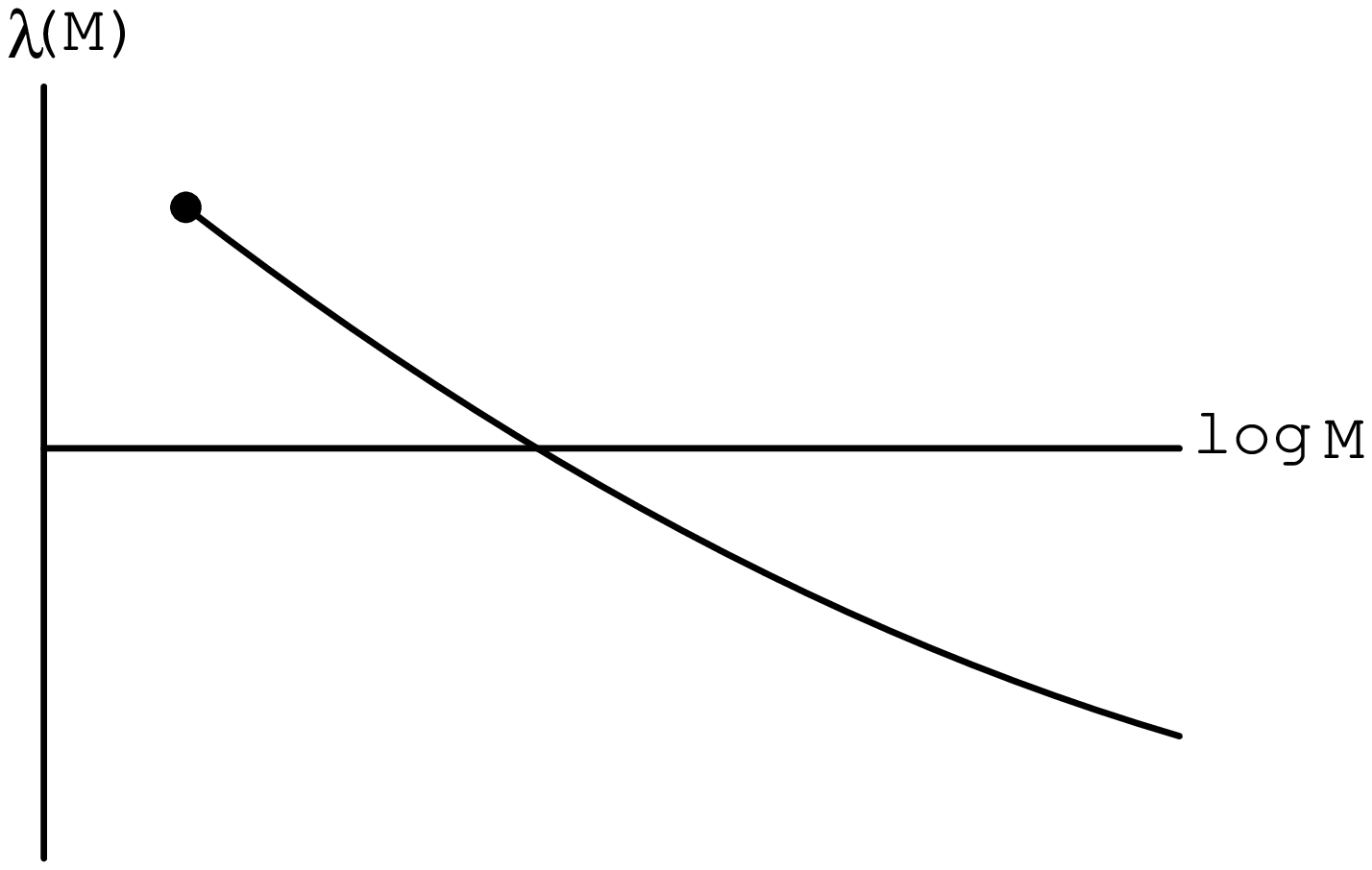}{60}

\PSfig\figb{The effective potential (a) classically and (b) including
  radiative corrections from a heavy top.}{4.5cm}{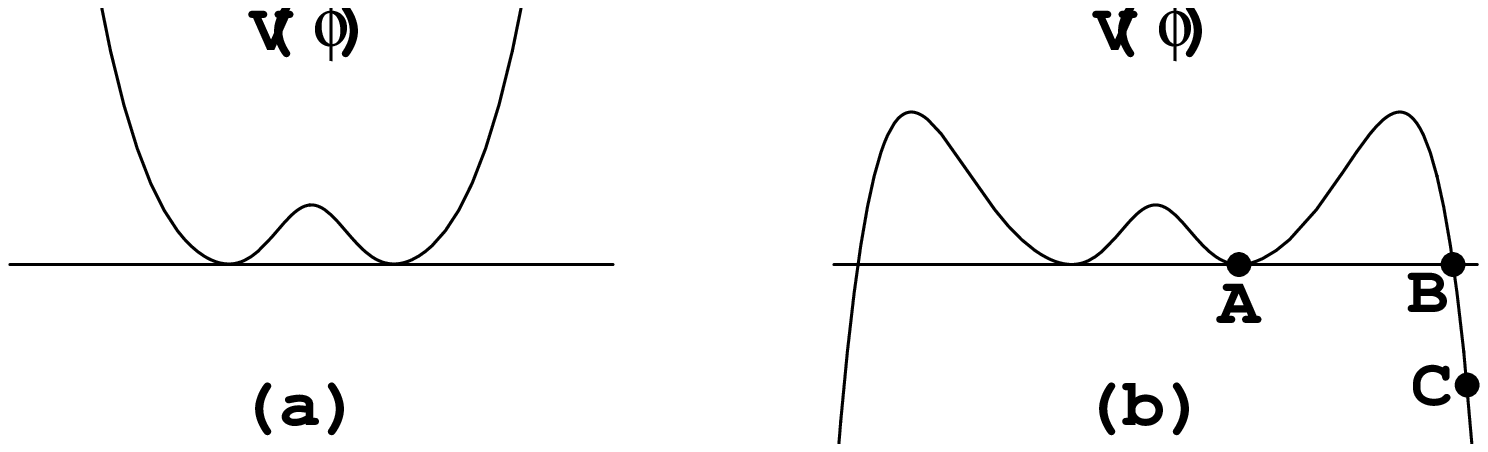}{80}

Now consider the classical Higgs potential shown in \figb a:
\eqn\eqVcl{
  \Vcl(\phi) = -{1\over2}\mu^2\phi^2 + {1\over4}\lambda\phi^4 .
}
When radiative corrections are included, one finds that the dominant
effect is to replace coupling constants by running coupling constants
evaluated at the scale of $\phi$ itself:
\eqn\eqVeff{
  \Veff(\phi) \approx - {1\over2} \mu^2(\bar\phi) \bar\phi^2
                      + {1\over4} \lambda(\bar\phi) \bar\phi^4 .
}
(More specifically, this is the result of summing leading-logarithms
to all orders for the effective potential.)
$\bar\phi$ is $\phi$ rescaled by its anomalous dimension.
But I shall not keep track of the distinction between $\phi$ and
$\bar\phi$; see Ref.~\av\ for details.

For large top mass, \figa\ and eq.~\eqVeff\ imply that effective potential
has the qualitative form of \figb b.  Our vacuum is only {\it meta}-stable,
and the potential is unstable
at large $\phi$!\refmark{\everyone-\twoloop}
Because the coupling
$\lambda$ runs logarithmically, the scale $\phi_{\rm B}$ at which the
potential becomes unstable in \figb b will in general be exponentially
large compared to the scale $\phi_A$ of the electroweak vacuum.

\PSfig\figc{Vacuum stability as a function of Higgs and top mass.
  Each line corresponds to a choice of the scale $\Lambda$ of physics
  beyond the Minimal Standard Model, and the solid line corresponds to
  choosing the Plank Scale.  Considering only $\phi<\Lambda$, our vacuum
  is absolutely stable above, and only meta-stable below, the
  corresponding line.}{11cm}{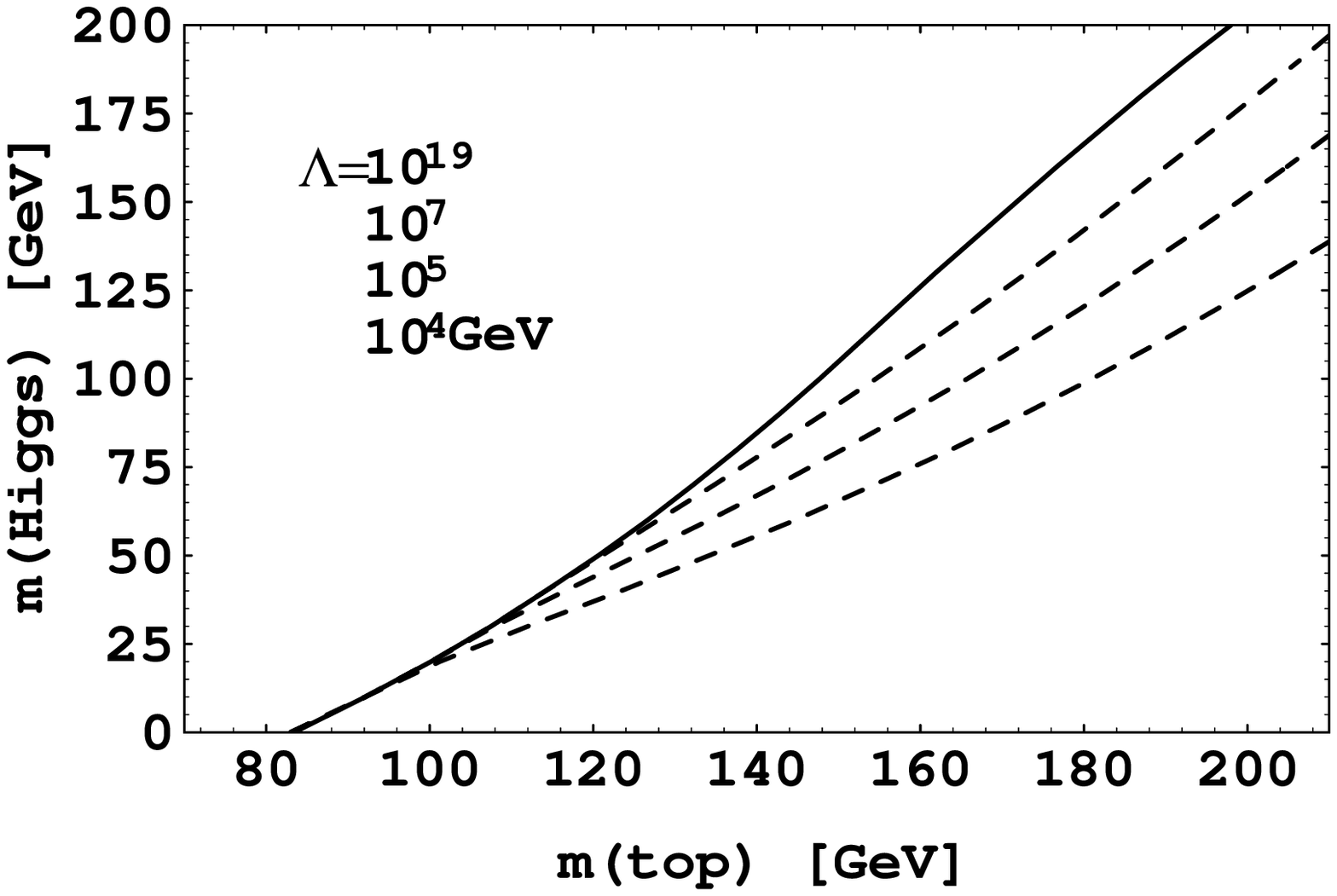}{65}

Fig.~\xfig\figc\ shows for which values of top and Higgs masses our
vacuum is unstable.  I treat the Minimal Standard Model as a good
effective theory up to some scale $\Lambda$.  The determination of
vacuum stability then depends on whether or not the instability
B of \figb b manifests, if it manifests at all, for $\phi<\Lambda$.
If so, then our vacuum is definitely unstable.  If not, then it is stable
within the assumed range of validity of the Minimal Standard Model, and any
instability could only arise from the unspecified, new physics above
$\Lambda$.

\PSfig\figd{The qualitative relationship between the potential energy
  and the radius of a bubble of unstable phase.}{9.8cm}{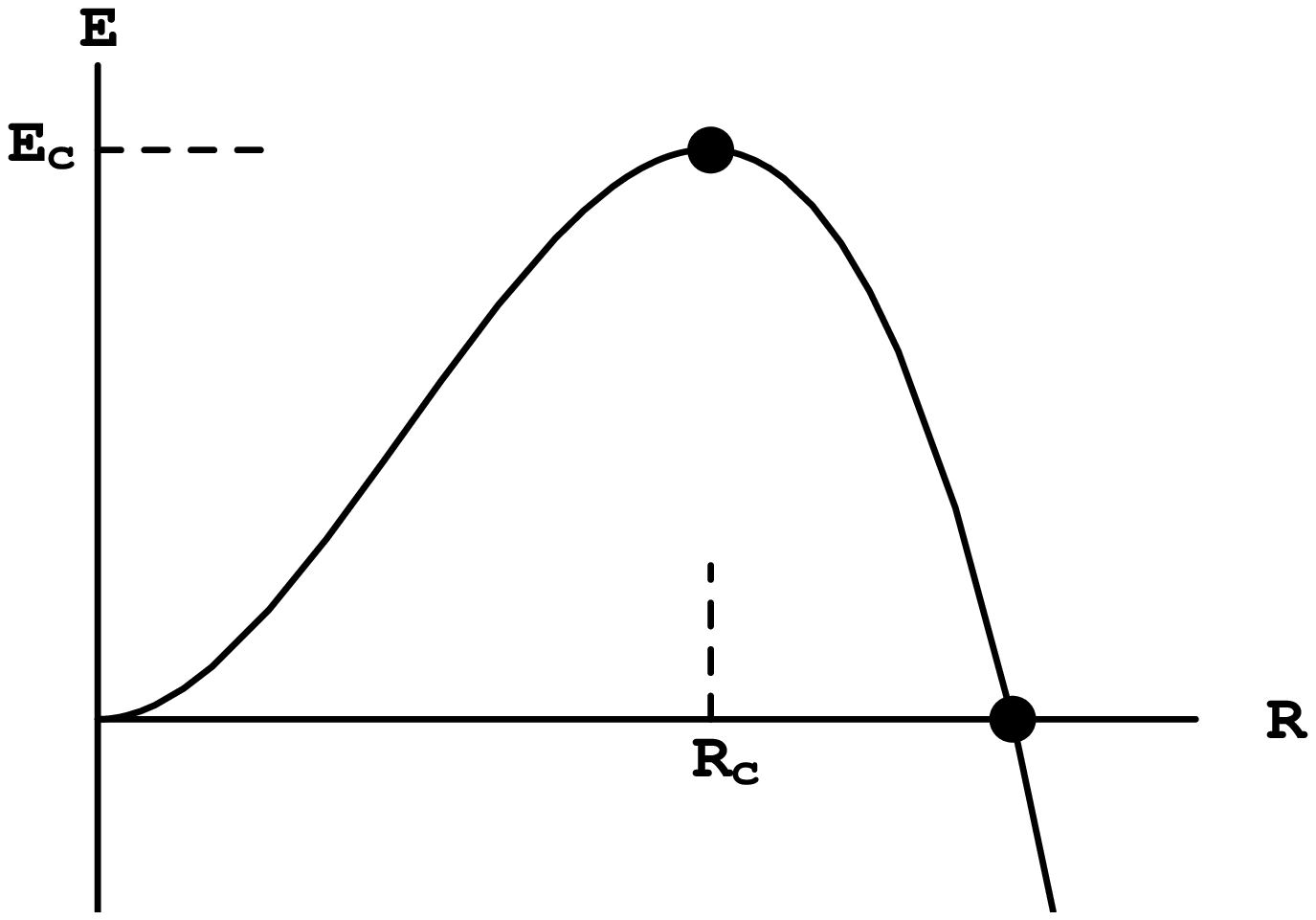}{60}

Vacuum decay is initiated by localized, random fluctuations---either
quantum or thermal---of the scalar field $\phi$.
Imagine a localized region of space of radius $R$ that probes the
instability of the potential of \figb b: the scalar field is of
order $\phi_{\rm C}$ inside $R$ and equals the meta-stable vacuum
$\phi_{\rm A}$ outside.  This ``bubble'' of unstable phase wants
to grow because the inside has lower potential energy than the
outside, which means that the total potential energy of the
bubble has a negative contribution which grows as -(volume).
On the other hand, the bubble wants to shrink because of
surface tension; there is a positive contribution to the total potential
energy [arising from the $(\nabla\phi)^2$ term] which grows with
$R$.  The potential energy of a bubble is shown qualitatively as
a function of radius $R$ in \figd.  (This picture artificially assumes the
order of magnitude of $\phi$ inside the bubble is held fixed.
See Ref.~\arnold\ for a qualitative picture of the more general case.)
Bubbles smaller than the critical bubble size $\Rc$ will collapse;
larger bubbles will grow, eventually sucking the entire universe
into the unstable phase.  (In most cases, the $\phi$ field will
continue rolling down the unstable potential until it reaches the
cut-off $\phi \sim \Lambda$, at which point its fate depends on what new
physics exists beyond the Minimal Standard Model.)

\newsec{Instability at Zero Temperature}

At zero energy, the only way our vacuum can decay is by quantum tunneling
beneath the barrier in \figd.  The rate for quantum tunneling through
a barrier is exponentially small, and the exponent is given by the
uncertainty principle.  If $\Delta E$ is the amount one must temporarily
violate energy conservation while tunneling beneath the barrier, and
$\Delta t$ is the amount of time one spends doing it, then very roughly
speaking
\eqn\Rtun{
  {\rm amplitude} \sim e^{-(\Delta E)(\Delta t)} \sim e^{-\SE} ,
}
where $\SE$ is the Euclidean action of the tunneling process.
In practice, one solves for the classical Euclidean ``bounce'' solution
that represents vacuum decay in the field theory problem and
evaluates its Euclidean action to get the amplitude $\exp(-\SE)$
for vacuum decay.  In the case at hand, this was done numerically
many years ago by Flores and Sher.\refmark\flores

\PSfig\fige{The dashed line is the $\Lambda=10^{19}$ GeV line from
  \figc\ and divides vacuum stability from meta-stability.  If only
  zero-temperature tunneling is considered, the lifetime of the vacuum
  exceeds 10 billion years to the left of the dot-dash line.
  The solid line shows the improved bound obtained by requiring the
  vacuum to be sufficiently stable in the hot early Universe.
  The dotted line is the related but less stringent bound from
  ref.~\greg.}{10.8cm}{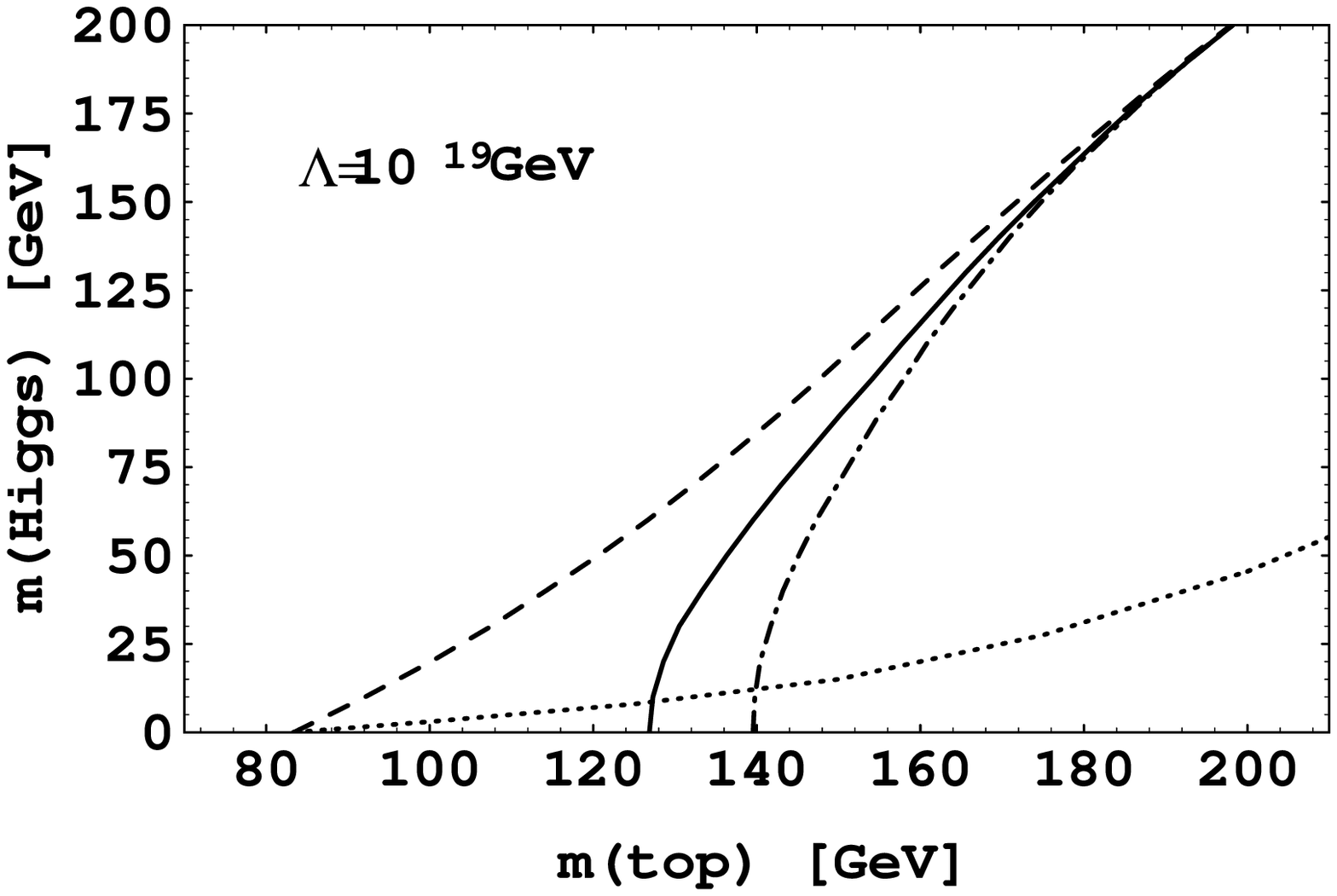}{65}

The observation that we exist does not require that our vacuum
be absolutely stable, but instead only requires that the lifetime
of our vacuum exceed the present age of the Universe.  Intriguingly,
there is a significant slice of parameter space for which our vacuum
is ultimately unstable but has a lifetime exceeding 10 billion years.
The dashed and dot-dash lines of Fig.~\xfig\fige\ delineate
this slice for the case $\Lambda=10^{19}$ GeV.
This computation was first performed by Flores and Sher.

There is a simple toy model for the tunneling problem\refmark\arnold\
which can be solved analytically and which gives an excellent approximation
(to within 1\% or so) to the full numerical results.
The effective potential \eqVeff\ only becomes unstable at exponentially
large values of $\phi$.  So to good approximation I may drop the
quadratic term in favor of the quartic one.  Since $\lambda$ only
runs logarithmically with $\phi$, let me approximate it in the
unstable region by a negative constant $-\kappa$.  The toy model is then
defined by the potential
\eqn\Vtoy{
  V(\phi) = -{1\over4} \kappa \phi^4.
}
The meta-stable vacuum of this model is at $\phi=0$.  At first glance
this may seem absurd because $V(\phi)$ above has no barrier against
vacuum decay.  But remember that the potential energy is
$(\nabla\phi)^2 + V(\phi)$ rather than just $V(\phi)$.  As a
result of the gradient term, $\phi=0$ is meta-stable against
any {\it local} perturbations.

To estimate the vacuum decay amplitude in this toy model, rescale
$\phi$ to factor the coupling constant out in front of the action:
\eqn\Skappa{
  \SE \rightarrow {1\over\kappa} \int\d^4x
  \left[ (\partial\tilde\phi)^2 - {1\over4}\tilde\phi^4 \right] .
}
So any non-trivial classical Euclidean solution has $\SE \sim 1/\kappa$.
This model is in fact well-known, and the relevant solution is
the Fubini instanton (or Lipaton):
\eqn\Sfubini{
  \SE = {8\pi^2\over3\kappa}, \qquad
  \phi(r) = \sqrt{2\over\kappa} {2R\over(\kappa^2+R^2)}
}
where the size scale $R$ of the solution is arbitrary because of the
scale invariance of \Skappa.  The rate of false vacuum decay in
this model then has exponential dependence
\eqn\ToyRate{
  {\rm rate} \sim e^{-8\pi^2/3\kappa} .
}

To return to the Minimal Standard Model, we only need to realize
that the scale invariance of the toy model is slightly broken by
the running of the coupling constants.  Tunneling will take place
at the scale which maximizes the rate \ToyRate, and so
\eqn\MSMrate{
  {\rm rate} \sim \max_{\lambda(\phi)<0}
  \left[ e^{-8\pi^2/3|\lambda(\phi)|} \right] .
}
This gives excellent agreement with numerical results.

\newsec{Instability at High Temperature}

At the high temperatures of the Early Universe, the situation is
completely different.  Return to the simple picture of the bubble
energy in \figd\ and consider temperatures large compared to the
energy barrier $\Ec$ for vacuum decay.  At such temperatures,
energy states with $E > \Ec$ will be thermally excited, and the
vacuum can then decay {\it classically}, without any exponential
suppression associated with quantum tunneling.  For more modest
temperatures, the probability of having enough energy to cross
the barrier classically is given by a simple Maxwell-Boltzmann
factor:
\eqn\maxwell{
  {\rm rate} \sim e^{-\beta\Ec} ,
}
where $\beta$ is the inverse temperature.

The critical energy
$\Ec$ is found by looking for a static, unstable solution to the
classical field equations of motion that corresponds to resting
precariously atop the barrier in \figd.  (In various contexts,
such solutions are called ``sphalerons'' or ``O(3) bounces.'')
Once the solution is found, $\Ec$ is identified as its classical
energy.  I shall discuss such solutions in the
Minimal Standard Model in a moment, but first I want to preview
the results.  The solid line in \fige\ shows the bound
that may be derived\refmark\av by requiring that our vacuum was stable enough
to persist through the hot temperatures of the early Universe.  Note
that it is {\it stronger} than the bound we derived earlier but
not {\it so} strong as to
require that our vacuum be absolutely stable.  This is the strongest
bound that has been derived from considerations of vacuum stability.
The dotted line shows an earlier bound by Anderson\refmark\greg\
that was based on considering temperatures only up to the critical
temperature of symmetry restoration in electroweak theory; our stronger
bounds come from investigating temperatures all the way up to the
assumed scale $\Lambda$ of new physics.

At first glance, these results may not make much sense.  If one looks
again at \figd, it appears that an unstable vacuum could {\it never}
survive the early Universe since, at some point in time, the
temperature must have been greater than the barrier energy $\Ec$.
In field theory, however, it happens that $\Ec$ is itself effectively
a function of $T$, and so the matter is more subtle.
At high temperature, the Higgs receives a thermal contribution to
its mass of order $g^2 T^2$ that is analogous to the Debye or plasmon
masses of the photon in a hot plasma:
\eqn\VT{
  \Veff(\phi,T) \approx \Veff(\phi,0) + {1\over2} g^2 T^2 \phi^2 ,
}
where we are adopting $g^2$ as a short-hand for a particular linear
combination of squared coupling constants.  In the Minimal Standard
Model,
\eqn\gdef{
  g^2 \rightarrow {1\over12}\left({3\over4}g_1^2 + {9\over4}g_2^2
      + 2\gy^2 + 6\lambda \right).
}
This $g^2 T^2 \phi^2$ term is the standard effect responsible for
symmetry restoration at high-temperature since it replaces
$-\mu^2$ in the classical potential \eqVcl\ by $-\mu^2+g^2 T^2$,
which turns positive for large $T$.

\PSfig\figg{The form of the effective potential relevant to the
  Linde-Weinberg bound (a) at zero temperature, and (b) at
  high temperature.}{4.5cm}{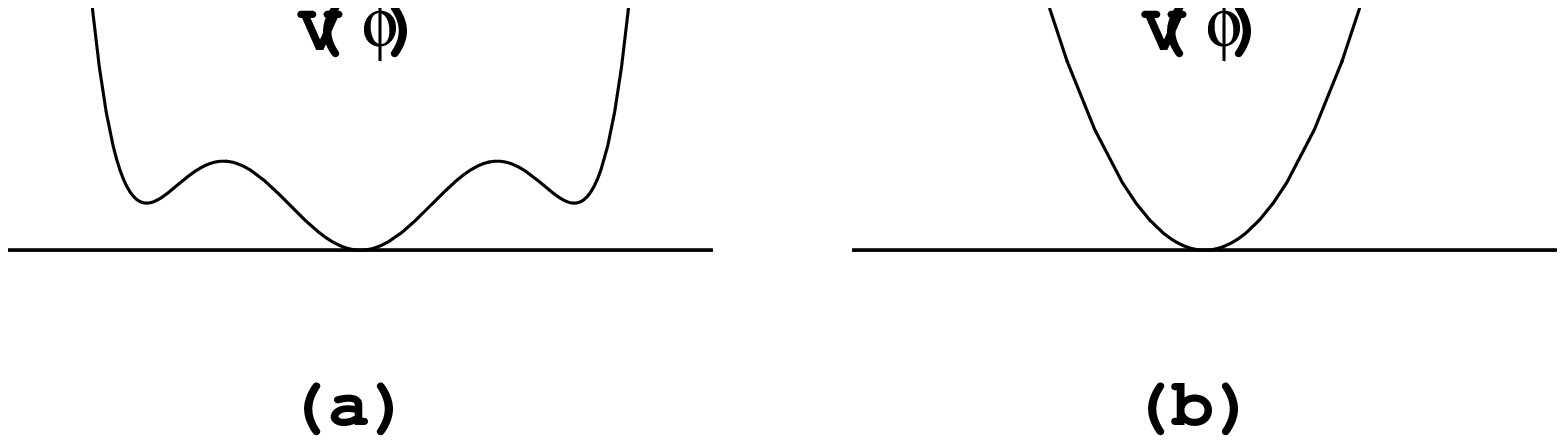}{80}

The thermal mass of the Higgs has an important impact in a problem similar
to the one considered in this talk: the Linde-Weinberg bound.  For
{\it small} $\mh$ and $\mt$ (rather than large $\mt$), the effective
potential is qualitatively of the form \figg a at zero temperature.
Our vacuum is again unstable but this time is unstable against decay
to $\phi\rightarrow 0$ rather than $\phi\rightarrow\infty$.  Once again,
there is a range of parameter space for which the lifetime of our
vacuum exceeds 10 billion years.  However, the thermal
mass term in \VT\ favors smaller $\phi$ over larger $\phi$.  As one
increases the temperature, the energy barriers between the vacua get
smaller and eventually disappear, so that the effective potential at
high temperatures looks like \figg b and has only the vacuum at $\phi=0$.
So, in the early Universe, the average value of $\phi$ will be zero.
By the time the Universe cools down enough for the non-zero meta-stable
vacua to appear, it is too late!  There is no way that the entire Universe
will jump into the false vacuum state.  As a result, the Higgs and top
masses that give rise to \figg a may all be ruled out.

In our case, however, the situation is reversed.  The barriers between
A and C in \figb b will {\it grow} with temperature.  (At the same time,
the potential will turn upward at the origin so that the false vacuum
$A$ moves to $\phi=0$.)  Thus, increasing the temperature involves a
trade-off for vacuum decay: (1) there is more energy available to thermally
excite transitions across the barrier, but (2) the barrier is higher.
If the universe started out in the false vacuum (which
depends on the nature of the new physics above $\Lambda$), then it
is not clear whether or not it would have decayed.
To resolve which effect wins, one must examine the critical energy
$\Ec$ for vacuum decay in more detail.

So let us return to the toy model of the previous section and consider its
high-temperature limit:
\eqn\VtoyT{
  V(\phi) \rightarrow {1\over2} g^2 T^2\phi^2 - {1\over4}\kappa\phi^4 .
}
Now consider the static energy $\int {1\over2}(\nabla\phi)^2 + V$ and
scale out both the coupling $\kappa$ and the dimension $gT$:
\eqn\scaleE{
  E \rightarrow {gT\over\kappa} \int \d^3\tilde x \left[
   {1\over2} (\nabla \tilde\phi)^2 + {1\over2}\tilde\phi^2
    - {1\over4} \tilde\phi^4 \right] .
}
The energy of any non-trivial static solution will therefore be of
order $gT/\kappa$.  Finding the extremum of the integral
numerically,\refmark\brezin\ one finds
\eqn\toyEc{
  \Ec = 6.015\pi \cdot {gT/\kappa},
}
and so the vacuum decay rate is roughly
\eqn\toyrate{
  {\rm rate} \sim e^{-\beta \Ec(T)} \sim e^{-6.0\pi g/\kappa} .
}
This result is independent of temperature.

The effect of the
returning to the real problem in the Minimal Standard Model is
to again break scale invariance.  The rate is maximal at the temperature
which maximizes $-\lambda(T)$:
\eqn\Trate{
  \max_T \left[\rm rate\right] \sim \max_{\lambda(T) < 0} \left[
     e^{-6.0\pi g(T)/|\lambda(T)|} \right] .
}
Note that, for sufficiently small coupling constants, this should always beat
the rate from zero-temperature tunneling because the exponents are of
order $\beta E \sim g/|\lambda|$ in the thermal case
and $\SE \sim 1/|\lambda|$ in the zero-temperature tunneling case.

A discussion of the dimensional prefactors in \Trate\ may be
found in Ref.~\av.

\newsec{Conclusion}

\PSfig\figh{The bound from vacuum decay in the early Universe shown
  for several values of the cut-off $\Lambda$.}{10.8cm}{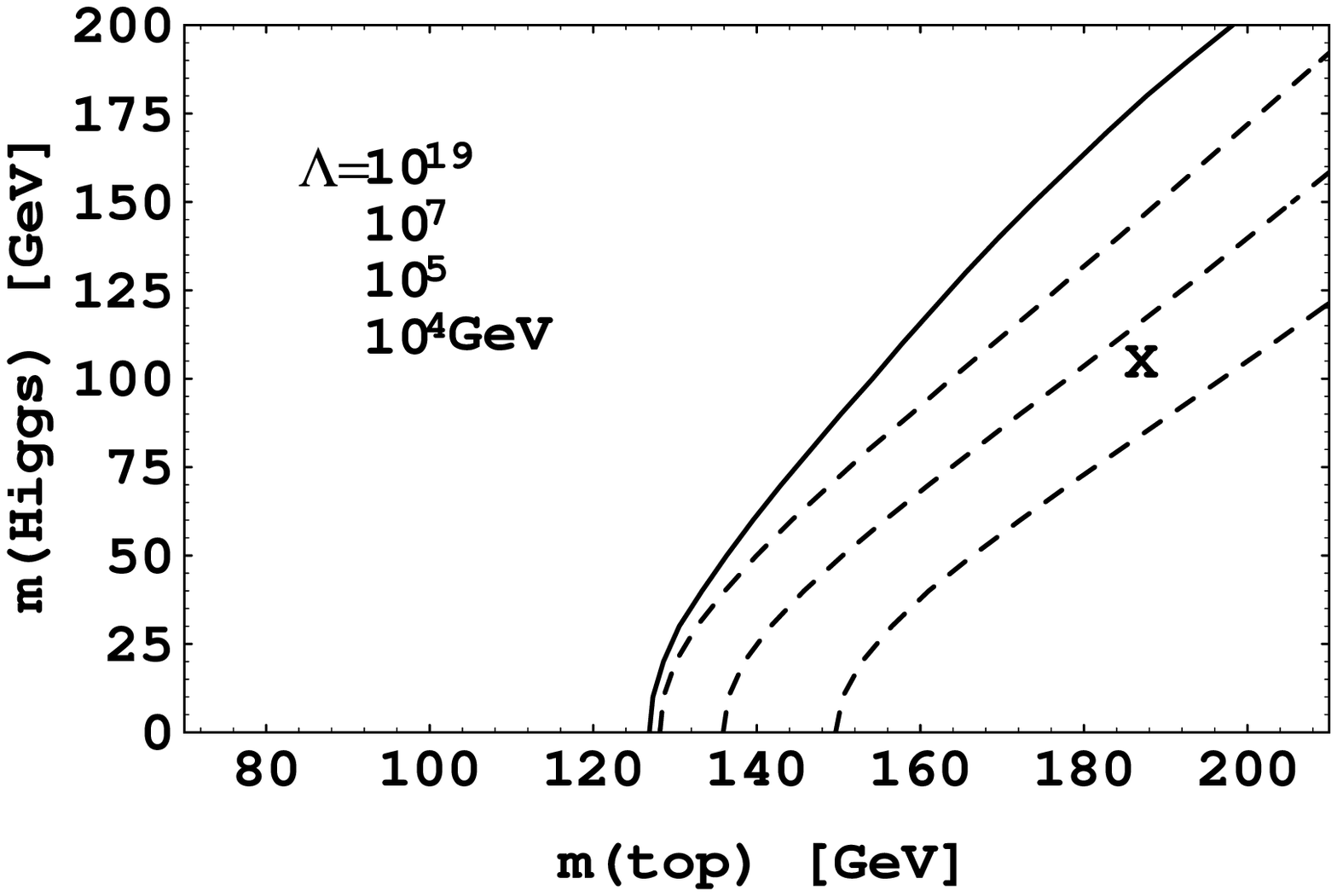}{65}

The result of applying \Trate\ to the Minimal Standard Model is shown
in \figh\ for a variety of choices of scale $\Lambda$ up to which
the Minimal Standard Model is posited to be a good effective theory.
What is all of this good for?  Suppose that sometime in the next decade
we happen to find the top and a candidate minimal Higgs at the point marked X.
We shall then know that there must be new physics below 100 TeV.
Alternatively, a sexier possibility is that we may someday discover that
the Minimal Standard Model is in fact a good effective theory up to
10 TeV or so and find the top and Higgs in a region of \figc\ that is
unstable for $\Lambda = 10^4$ GeV.  We shall then {\it know} that the
Universe as we know it is doomed and that our vacuum will decay someday
billions of years hence.

The reader who is interested in much more speculative bounds, based on
the controversial idea that weak interactions may necessarily become
strong at high energy, should examine Refs.~\ellis\ and \hsu.

\bigskip
\footatend\immediate\closeout\rfile\writestoppt
\centerline{{\bf References}}\bigskip{\frenchspacing%
\parindent=20pt\escapechar=` \input \jobname.refs\vfill\eject}\nonfrenchspacing

\if\preprint0 \listfigs \fi

\end